\documentclass[sigconf]{acmart}
\AtBeginDocument{%
  \providecommand\BibTeX{{%
    \normalfont B\kern-0.5em{\scshape i\kern-0.25em b}\kern-0.8em\TeX}}}

\setcopyright{acmcopyright}
\copyrightyear{2024}
\acmYear{2024}
\acmDOI{3703412.3703416}

\acmISBN{979-8-4007-1161-9}

\usepackage{microtype}
\usepackage{hyperref}
\usepackage{hyperxmp}
\usepackage{url}
\usepackage{booktabs}

\usepackage{graphicx}
\usepackage[normalem]{ulem}
\useunder{\uline}{\ul}{}
\usepackage{vcell}
\usepackage{multirow}

\copyrightyear{2024}
\acmYear{2024}
\setcopyright{rightsretained}
\acmConference[AIMLSystems 2024]{4th International Conference on AI-ML Systems}{October 08--11, 2024}{Baton Rouge, LA, USA}
\acmBooktitle{4th International Conference on AI-ML Systems (AIMLSystems 2024), October 08--11, 2024, Baton Rouge, LA, USA}
\acmPrice{}
\acmDOI{10.1145/3703412.3703416}
\acmISBN{979-8-4007-1161-9/24/10}

\begin{document}

\title{BudgetMLAgent: A Cost-Effective LLM Multi-Agent system for Automating Machine Learning Tasks}


\author{Shubham Gandhi}
\affiliation{%
  \institution{TCS Research}
  \city{Pune}
  \country{India}
  }
\email{gandhi.shubham@tcs.com}

\author{Manasi Patwardhan}
\affiliation{%
  \institution{TCS Research}
  \city{Pune}
  \country{India}
  }
\email{manasi.patwardhan@tcs.com}

\author{Lovekesh Vig}
\affiliation{%
  \institution{TCS Research}
  \city{Delhi}
  \country{India}
  }
\email{lovekesh.vig@tcs.com}

\author{Gautam Shroff}
\affiliation{%
  \institution{TCS Research}
  \city{Delhi}
  \country{India}
  }
\email{gautam.shroff@tcs.com}

\renewcommand{\shortauthors}{Gandhi, et al.}

\begin{abstract}
Large Language Models (LLMs) excel in diverse applications including generation of code snippets, but often struggle with generating code for complex Machine Learning (ML) tasks. Although existing LLM single-agent based systems give varying performance depending on the task complexity, they purely rely on larger and expensive models such as GPT-4. Our investigation reveals that no-cost and low-cost models such as Gemini-Pro, Mixtral and CodeLlama perform far worse than GPT-4 in a single-agent setting.  With the motivation of developing a cost-efficient LLM based solution for solving ML tasks, we propose an LLM Multi-Agent based system which leverages combination of experts using profiling, efficient retrieval of past observations, LLM cascades, and ask-the-expert calls. Through empirical analysis on ML engineering tasks in the MLAgentBench benchmark, we demonstrate the effectiveness of our system, using no-cost models, namely Gemini as the base LLM, paired with GPT-4 in cascade and expert to serve occasional ask-the-expert calls for planning. With 94.2\%  reduction in the cost (from \$0.931 per run cost averaged over all tasks for GPT-4 single agent system to \$0.054), our system is able to yield better average success rate of 32.95\% as compared to GPT-4 single-agent system yielding 22.72\%  success rate averaged over all the tasks of MLAgentBench. 
\end{abstract}




    \maketitle

\section{Introduction}

Although recent advances have shown that Large Language Models (LLMs) are adept at handling a vast array of applications ranging from natural language \citep{fang2024large, huang2023reasoning, zhu2023multilingual, yi2024survey} to code-related tasks \citep{zheng2024survey, zan-etal-2023-large, zhang2024unifying}, this capability does not often translate to more complicated and nuanced tasks \citep{yeadon2024comparison}. Most code-related efforts involving LLMs \citep{guo2024deepseekcoder, huang2024agentcoder, zhong2024ldb} are based on tasks such as HumanEval \citep{chen2021evaluating} and MBXP \citep{athiwaratkun2023multilingual}, that have a relatively easier level of complexity that is far from what is experienced by data scientists. However, real-world engineering challenges demand nuanced problem-solving and intricate planning, often involving multiple rounds of strategizing, experimentation, and recalibration. LLM agent systems excel in simulating this iterative process, since they comprise of an environment containing code files, description files and data files and a pre-defined action space allowing interaction with the environment. This demonstrates their capability to address intricate engineering challenges effectively. \citep{zhang2024codeagent}.

Transitioning to codifying Machine Learning (ML) applications brings its own challenges since they often involve training models on datasets, tuning hyperparameters, devising ways to improve performance, etc. These applications are not straightforward and require a deep understanding of the underlying algorithms and techniques along with specific libraries used for implementation of plans. Although there exist AutoML-based approaches for automating such tasks \citep{He_2021, SALEHIN202452}, these offer limited flexibility since they typically operate within predefined constraints and search spaces in the form of possible configurations of architectures and/ or hyper-parameters, which may limit their ability to explore solutions out-of-distribution of the search space. While works such as ChatDev \citep{qian2023communicative} and MetaGPT \citep{hong2023metagpt} have explored the capabilities of LLM Agents in a software development environment, there is a notable scarcity of research on utilizing LLM Agents for solving ML tasks. 

Recent works like MLCopilot \citep{zhang2024mlcopilotunleashingpowerlarge} introduce an assistant for solving ML tasks. However, these architectures are limited in the types of problems they can address and must strictly follow task description formats that do not align with real-world scenarios. Additionally, such assistants only suggest solutions, leaving the actual burden of implementation to the user.
To the best of our knowledge, MLAgentBench \citep{huang2023benchmarking} is the only significant benchmark addressing ML problem solving capabilities of LLM Agents directly dealing with code. 

Although they get good performance on some tasks in their benchmark, they focus on single-agent systems using expensive LLMs such as GPT-4, which  costs approximately \$0.52-\$2.9 per run, depending on the task. For the experiments they conduct, they go for 8 runs per task for 15+ tasks, leading to very high experimental cost of approximately \$200+.
With such larger models becoming increasingly expensive to use, there is a natural incentive to develop no-cost or low-cost systems using smaller, open-source models and making them equally capable for niche tasks. However, existing agent creation frameworks like AutoGen \cite{wu2023autogen} do not prioritize cost-reduction. Replacing single-agent systems using expensive LLMs with single-agent smaller, open-source LLMs may not serve the purpose.  Our initial experiments with replacing all LLM calls in \citet{huang2023benchmarking} for auto-generating codes for ML tasks, with no or low-cost LLMs, namely, Gemini-Pro \citep{geminiteam2023gemini}\footnote{gemini-pro 1.0 API from \url{https://ai.google.dev/tutorials/python_quickstart}. The rate limit for the free or no-cost version is sufficient for conducting our experiments, however, we also include costs for a no-cost version with pay-as-you-go pricing}, CodeLlama \citep{rozière2024code}\footnote{\url{https://huggingface.co/codellama/CodeLlama-34b-Instruct-hf}} and Mixtral\citep{jiang2024mixtral}\footnote{Mixtral-8x7B-v0.1  \url{https://huggingface.co/mistralai/Mixtral-8x7B-v0.1}}, yield very poor results for all of the tasks in a single-agent setting. 

In real-world setting any complicated tasks are rarely tackled by a single individual alone, especially when all the individuals do not possess the required expertise to perform the task. Instead, teams of engineers collaborate, with each member having a unique role (\textit{persona}) and contributing unique expertise and skills to achieve the target with collective efforts. Past works on LLM agents have simulated this real-world setting by designing multi-agent frameworks \citep{li2024Agents, shen2024small}, combining LLM experts \citep{wang2023fusing, ding2024hybrid} and defining cascades \citep{chen2023frugalgpt, yue2024large, zhang2023ecoassistant} for tasks such as code generation, reasoning, question answering, etc. Cascades refer to the chaining of LLMs in a progressive fashion, where, a weaker LLM is invoked first and if the response is not satisfactory then stronger LLMs are invoked. However, to the best of our knowledge multi-agent frameworks with open-source LLMs as agents have not be explored for engineering of ML tasks.

In this paper, we address the gap of utilizing LLMs for solving ML tasks by proposing a system that leverages - (i) Multi-LLM Agents as a combination of experts using profiling, (ii) LLM Cascades, (iii) Efficient retrieval of relevant past observations, and (iv) our novel occasional ask-the-expert calls to GPT-4 \footnote{gpt-4-0125-preview \url{https://platform.openai.com/docs/models/gpt-4-and-gpt-4-turbo}} for planning. Our approach aims to bridge the divide between capabilities of cheaper LLMs and the requirements of complex ML tasks, offering a more cost-efficient and scalable solution. 
Through empirical analysis, we validate the following claims:
\begin{itemize}
\item  Our best performing multi-agent system using no-cost or low-cost versions of Gemini-Pro as the base LLM, is able to perform tasks at a fraction of the cost (on an average average \$0.054 for no-cost and \$0.120 for low-cost version per run per ML task in MLAgentBench Dataset) as compared to benchmarked  single-agent GPT4 system presented in \citet{huang2023benchmarking} (on an average \$0.931 per run per task)
\item With 94.2\% and 87.1\% cost reduction for the no-cost and low-cost Gemini-Pro versions, our best performing multi-agent system is able to yield better success rate of 32.95\% averaged for all the tasks in MLAgentBench as compared to the GPT4 based single-agent system yielding 22.72\% average success rate for all tasks.
\item Our best performing multi-agent system is able to achieve equal or better performance for 45.45\% of tasks when compared to the GPT4-based Single-Agent system in \citet{huang2023benchmarking}, whereas it yields comparable performance for other tasks
\end{itemize}
\section{MLAgentBench Dataset}\label{sec:mlagentbench}

MLAgentBench \citep{huang2023benchmarking} is a dataset designed for evaluating LLM Agents for Machine Learning (ML) tasks. ML tasks defined within the MLAgentBench dataset are specified with clarity, providing a concise description of the desired objective, evaluation metric, and submission guidelines. These task descriptions are diverse and not restricted to a particular format, allowing for a broad range of task definitions.
For example, tasks involve improving model accuracy on a given dataset or optimizing a specific performance metric. The dataset also provides necessary files containing training and testing data, along with detailed descriptions of the data and metrics. Starter code, implemented across diverse ML frameworks like PyTorch, TensorFlow, JAX, and Keras, is provided to assist agents in getting started. While some tasks (cifar10, ogbn-arxiv, etc) offer baseline implementations for comparison, others (imdb, house-price, etc) require agents to code models from scratch based on the provided specifications and dataset files.

In the MLAgentBench framework, each task represents an environment where agents interact by performing actions and receiving observations. The benchmark offers a set of primitive low-level actions, including file system operations (For example, list files, read, write, append, copy, etc), executing Python scripts, and declaring final answers. Additionally, there also exist high-level actions such as understanding a file, reflection (looking over past steps and contemplating based on the given description of what to reflect on), inspecting a segment of a file and editing a script (or a script segment). High-level actions may call some low-level actions or LLMs internally (for example understand file action might result in file contents being passed to an LLM and asking it to understand the contents.). Each action is accompanied by comprehensive documentation, specifying its name, description, usage guidelines, expected return values, and implementation. These actions enable LLM agent to manipulate files, execute scripts, and declare final outcomes within the task environment, facilitating iterative problem-solving and evaluation. 

We consider a subset of MLAgentBench dataset. We do not take two tasks in to consideration, viz. fathomnet \citep{fathomnet-out-of-sample-detection} and identify-contrails \citep{google-research-identify-contrails-reduce-global-warming}, due to our compute restrictions which can not handle the extremely large size of the datasets corresponding to these tasks. We also drop the LLM Tools tasks from MLAgentBench, as it does not have any comparable results for evaluating our system. Thus, we take into consideration the following tasks: (i) Canonical tasks sich as cifar10 \citep{Krizhevsky2009LearningML}, imdb \citep{maas-etal-2011-learning} and ogbn-arxiv \citep{hu2021open}), (ii) Classic Kaggle tasks such as house-price \citep{house-prices-advanced-regression-techniques} and spaceship-titanic \citep{spaceship-titanic}, (iii)  Kaggle Challenges such as parkinsons-disease \citep{amp-parkinsons-disease-progression-prediction} and feedback \citep{feedback-prize-english-language-learning})  (iv) Current Research  such as CLRS \citep{veličković2022clrs} and BabyLM \citep{warstadt2023papers}) and (v) Improve Code tasks  such as llama-inference and vectorization.
\section{BudgetMLAgent} \label{sec:budgetmlagent}

\begin{figure*}[!ht]
  \centering
  \includegraphics[width=\textwidth]{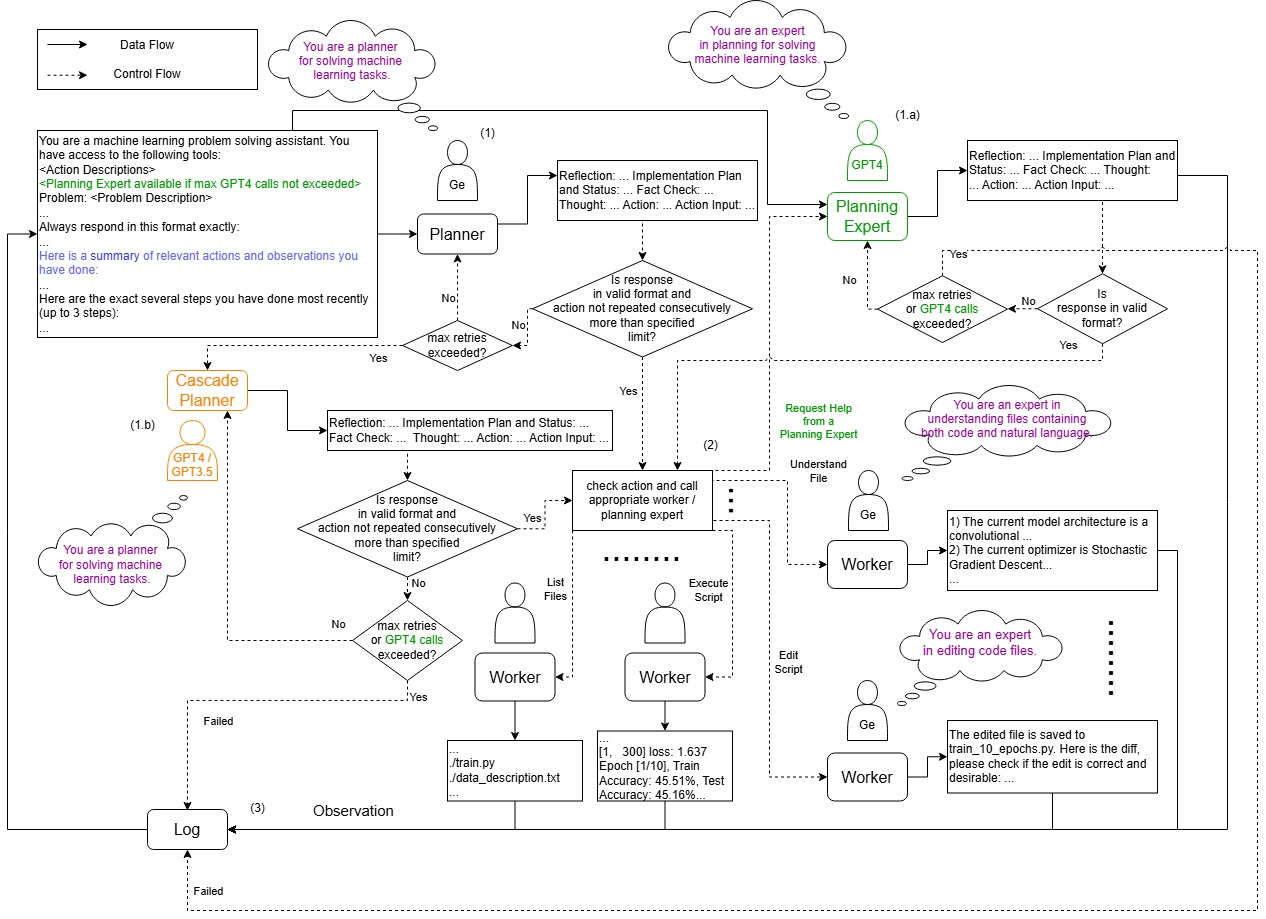}
  \caption{BudgetMLAgent defined in Section \ref{sec:budgetmlagent} with \color[HTML]{990099}{profiling}, \color[HTML]{FF8000}{cascades}, \color[HTML]{009900}{ask-the-expert lifelines} \textcolor{black}{and} \color[HTML]{9999FF}{retrieval from logs}. \textcolor{black}{Note that GPT4 cascade and planning expert calls both count towards the} \color[HTML]{009900}{max GPT4 calls limit. }\textcolor{black}{Ge: Gemini Pro (no-cost LLM Agent).} }
  \label{fig:budgetmlagent}
\end{figure*}

Our proposed system, BudgetMLAgent is an LLM Multi-Agent system that primarily uses no-cost LLMs and incorporates several enhancements viz. (i) LLM Profiling, (ii) LLM Cascades and (iii) Ask-the-expert lifelines. 

\subsection{Multi-Agent LLM Profiling}\label{sec:multi-llm}

Our system capitalizes on the groundwork laid  by MLAgentBench \citep{huang2023benchmarking} that provides a straightforward single LLM Agent based solution for the tasks. It operates through an organized prompt-response based interaction system by an agent that uses a set of available actions to interact with the environment. Through carefully structured prompts, they aim to ensure clarity and precision in conveying task descriptions, available tools (possible set-of actions), and most recent steps taken, to enhance the agent's decision-making process. To emphasize thoughtful decision-making during planning, the LLM is instructed to stick to a structured format for providing responses to the aforementioned structured prompts, including elements such as `Reflection' on understanding the prior observations, an updatable `Implementation Plan' and step-wise `Status', `Fact check' on if the objective statements from the Plan and Status guessed or directly confirmed and `Thought' on the action to be performed with justification and reasoning. This should be followed by the proposed `Action' for the next step along with the corresponding `Action Inputs' in JSON format. This structured response format is aimed at enhancing the agent's ability to engage in reflective thinking, better planning, and result verification. They also make use of a logging mechanism inspired by the memory stream paradigm \citep{park2023generative}, which enables efficient management of historical data rather than inundating the LLM with extensive historical context. By adopting this design, they ensure that the log file serves as a repository of relevant information that can be easily retrieved and updated by the agent using LLMs. Retrieval-enabled ($R$) runs refer to the ones having this functionality enabled. Thus, this retrieved information coupled with the recent actions and observations make up the historical context. The retrieved information acts as long-term memory whereas the recent actions and observations act as short-term memory.


\begin{table*}[t]
\centering
\resizebox{\textwidth}{!}{%
\begin{tabular}{@{}cll@{}}
\toprule
\multicolumn{1}{l}{\textbf{Type}} & \textbf{Agent} & \textbf{Profile} \\ \midrule
\multirow{2}{*}{\textbf{\begin{tabular}[c]{@{}c@{}}Planner\end{tabular} }} & Default planner & You are a planner for solving machine learning tasks. \\
 & Planning Expert & You are an expert in planning for solving machine learning tasks. \\ \midrule
\multirow{4}{*}{\textbf{\begin{tabular}[c]{@{}c@{}}High-level\\action worker\end{tabular} }} & Understand File & You are an expert in understanding files containing both code and natural language. \\
 & Edit Script (AI) & You are an expert in editing code files. \\
 & Reflection & You are an expert in reflecting on previous actions when solving a machine learning task. \\
 & Inspect Script Lines & NA \\ \midrule
\multirow{5}{*}{\textbf{\begin{tabular}[c]{@{}c@{}}Low-level\\action worker\end{tabular} }} & List Files & NA \\
 & Copy File & NA \\
 & Undo Edit Script & NA \\
 & Final Answer & NA \\
 & Execute Script & NA \\ \bottomrule
\end{tabular}%
}
\caption{Agents for callable actions and profiles given through system prompts to agents involving LLM calls. NA - no profile as these are programmatic agents and not LLM agents.}
\label{tab:system-prompts}
\end{table*}

We extend the above framework for a multi-agent LLM system. Multi-Agent LLM Profiling refers to the technique of combining the expertise of multiple agents using LLM Profiling, i.e. assigning each of the agents distinct personas. We characterize the multi-agent nature of our system by categorizing the agents into two specific classes - (i) A Planner (P) that utilizes the aforementioned agent structure to consider historical context and 'plan' the next action, and (ii) Workers (W${_i}$s) that execute the actions. In addition to a profile for the planner, we also include distinct personas for workers performing distinct actions that involve calls to LLMs such as \textit{Edit Script}, \textit{Understand File}, etc as seen in table \ref{tab:system-prompts}. Instead of having the default \textit{"You are a helpful AI assistant"} system prompt, we have distinct profiles for each action resulting in a system with agents specialized in distinct roles. These worker agents do not interact with each other and are instead invoked by the Planner P agent whenever it chooses to perform a corresponding action. Executing these actions may or may not involve an internal LLM call.


\subsection{LLM Cascade}
\label{sec:llm-cascade}
LLM Cascades refer to the technique of conditional invocation of sequentially connected LLMs (L$_1$, L$_2$, ... L$_k$). Here, we chain LLMs in a manner wherein $cost(L_1) < cost(L_2) ... < cost(L_k)$. Here the cost is represented by the latest pricing information of the corresponding models. 
A set of protocols are enforced to decide if the response by an LLM at a particular "cascade" is acceptable or not. If it is acceptable, then the response is used as is and if not, we move up the cascade to the next LLM. 
For example, the LLMs in cascade could be Gemini-Pro (a no-cost LLM) followed by GPT4 (an expensive LLM). If Gemini-Pro fails to generate an acceptable response before exhausting its maximum retries, then the system would invoke GPT4 for that step.
In our study, the protocols to move up the cascade are two-fold - (i) If the current LLM fails to generate a response that adheres to the specified format, even after maximum $m$ number of tries, or (ii) If the current LLM chooses an action that has already been repeated $r$ consecutive times in the past $r$ steps. 

\subsection{Ask-the-Expert Action}

To save on expenses, we primarily employ no-cost LLMs for both planning and action-based LLM calls. Preliminary investigations reveal that mostly planning is the  shortcoming of these LLMs and that their responses are of sufficient quality when it comes to other action-based LLM calls by workers. Therefore, we give the LLM with the Planner P persona, $l$ 'lifelines', where it can choose to call a larger, more expensive LLM with higher expertise,
in scenarios where it identifies that it is stuck at a step.
This choice is implemented in the form of an action, \textit{Request Help from a Planning Expert}, that the current `Planner' LLM agent can choose to take. Moreover, in practice, this upper cap on number of larger model calls (life-lines) also counts the calls made as a result of the cascade protocol mentioned in Section \ref{sec:llm-cascade}. During the course of a run, once the planner hits this upper cap, the \textit{Request Help from a Planning Expert} action is no longer included in the prompt when populating the actions available to call.

\section{Experimentation and Results}

We perform our experiments on the subset of tasks of  MLAgentBench dataset explained in Section \ref{sec:mlagentbench}.

\subsection{Metrics}

We evaluate our results by taking two metrics under consideration.

\subsubsection{Success Rate}

The success rate is the percentage (\%) of runs which are considered as successful. 
As per defined in \citet{huang2023benchmarking}, a run is considered to be successful if it achieves more than 10\% improvement at the last step over the average performance of the baseline in the starter code. Here the performance measure is task specific. For canonical tasks, classic kaggle, kaggle challenges and current research type tasks mentioned in Section \ref{sec:mlagentbench}, the prediction accuracy of the final submission.csv file is considered as the performance metric. For improve code type tasks, the improvement in the runtime of the code is considered as a success metric, whereas for CLRS, the saved final model checkpoints are evaluated for accuracy and improvement in accuracy is considered as success criteria.



\subsubsection{Cost}

If an LLM has a monetary cost associated with it, we compute the average cost in dollars (\$) per run based on number of tokens used for that model\footnote{ The pricing information as of March 2024}. For LLMs where APIs are available, this becomes $\$[($Cost\_per\_input\_token $*$ Num\_of\_input\_tokens$) + ($Cost\_per\_output\_token $*$ Num\_of\_output\_tokens$)]$. Claude1 V1.0 used as a single agent in \citet{huang2023benchmarking}  is discontinued and hence latest pricing details for this are unavailable. For approximating cost for Claude, we use pricing information for Claude-Instant \footnote{\url{https://www.anthropic.com/api}}.

\subsection{Models and hyperparameters}

We analyze multiple no-cost LLMs such as Gemini, CodeLlama and Mixtral as single agents to test their ML problem solving capabilities. We employ two configurations for our proposed multi-agent framework: (i) Gemini, which is best performing single agent LLM, with ChatGPT \footnote{gpt-3.5-turbo from \url{https://platform.openai.com/docs/models/gpt-3-5-turbo}} in cascade. (ii) Gemini with  GPT4 in cascade and `Ask-the-Expert' agent.   For our runs, we set all hyperparameters as per the implementation of \citet{huang2023benchmarking}.   Maximum number of actions is set to 30. Maximum number of recent actions to be included in context as the short-term memory is set to to 3. For runs involving cascades, we set the maximum number of retries allowed ($m$) to 3 for Gemini-Pro, 3 for ChatGPT and 1 for GPT4 in the interest of cost. For ask-the-expert GPT4 calls also, we set the maximum number of retries allowed to 1. The maximum number of times an action can be consecutively repeated ($r$) is set to 3. For planning-related LLM calls, the temperature is set to 0.2 and for internal action-related LLM calls, the temperature is set to 0.01, since we need some amount of diversity in the output for the former and more definitive responses for latter.
For runs involving ask-the-planning-expert calls, we use GPT4 as the planning expert and set the maximum number of calls to GPT4 (lifelines $l$) to 5. Note that this also includes the calls made to GPT4 in an LLM cascade. Since the exact monetary cost for GPT4 and Claude single agent runs from \citet{huang2023benchmarking} is not made available, we approximate these costs for comparison. We use the average token usage from other single agent runs, namely Gemini-Pro and CodeLlama, after multiplying with a factor of 0.809 to account for 19.1\% lesser token usage by GPT4 as suggested by \citet{huang2023benchmarking}. 

\subsection{Baselines}
We use following baselines: (i) Single Agent GPT4 in retrieval setting (G + R) (ii) Single Agent GPT4 with no retrieval (G) (iii) Single Agent Claude V1.0 in retrieval setting (C + R) (iv) Single Agent Claude V1.0  with no retrieval (C) ((i) to (iv) are from \citet{huang2023benchmarking}) (v) Single Agent Gemini Pro in retrieval setting (Ge + R) (vi) Single Agent Code Llama in retrieval setting (Co + R) (vii) Single Agent Mixtral in retrieval setting (Mx + R) ((v) to (vii) are our baselines using no-cost LLMs)

\begin{table*}[]
\centering
\renewcommand{\arraystretch}{1.2}
\resizebox{\textwidth}{!}{%
\begin{tabular}{@{}lllllllllll@{}}
\toprule
   &
  \multicolumn{6}{c}{\textbf{Single Agent}\citep{huang2023benchmarking}} &
  \multicolumn{2}{c}{\textbf{Profiling + Cascade}} &
  \multicolumn{2}{c}{\textbf{Profiling + Cascade + Expert}} \\ \cmidrule(lr){2-7} \cmidrule(lr){8-9} \cmidrule(lr){10-11}
\textbf{Task} &
  \multicolumn{1}{l}{\textbf{G + R}} &
  \multicolumn{1}{l}{\textbf{G}} &
  \multicolumn{1}{l}{\textbf{C + R}} &
  \multicolumn{1}{l}{\textbf{C}} &
  \multicolumn{1}{l}{\textbf{Ge + R}} &
  \multicolumn{1}{l}{\textbf{Ge}} &
  \multicolumn{1}{l}{\textbf{Ge + Ch + R}} &
  \multicolumn{1}{l}{\textbf{Ge + Ch}} &
  \multicolumn{1}{l}{\textbf{Ge + G + R}} &
  \multicolumn{1}{l}{\textbf{Ge + G}} \\ \midrule
\textbf{\begin{tabular}[c]{@{}l@{}}cifar10 \end{tabular}} &
  {\begin{tabular}[c]{@{}l@{}}25\\ (\$0.583)\end{tabular}} &
  {\begin{tabular}[c]{@{}l@{}}50\\ (\$0.44)\end{tabular}} &
  {\begin{tabular}[c]{@{}l@{}}8\\ (\$0.058)\end{tabular}} &
  {\begin{tabular}[c]{@{}l@{}}48\\ (\$0.044)\end{tabular}} &
  {\begin{tabular}[c]{@{}l@{}}12.5\\ (\$0)*\\ (\$0.016)\end{tabular}} &
  {\begin{tabular}[c]{@{}l@{}}0\\ (\$0)*\\ (\$0.044)\end{tabular}} &
  {\begin{tabular}[c]{@{}l@{}}62.5\\ (\$0)*\\ (\$0.025)\end{tabular}} &
  {\begin{tabular}[c]{@{}l@{}}25\\ (\$0)*\\ (\$0.093)\end{tabular}} &
  {\textbf{\begin{tabular}[c]{@{}l@{}}\underline{75}\\ \underline{(\$0.057)*}\\ \underline{(\$0.094)}\end{tabular}}} &
  {\begin{tabular}[c]{@{}l@{}}37.5\\ (\$0.06)*\\ (\$0.034)\end{tabular}} \\
\textbf{\begin{tabular}[c]{@{}l@{}}imdb\end{tabular}} &

  {\begin{tabular}[c]{@{}l@{}}12.5\\ (\$1.48)\end{tabular}} &
  {\textbf{\begin{tabular}[c]{@{}l@{}}\underline{25}\\ \underline{(\$0.919)}\end{tabular}}} &
  {\begin{tabular}[c]{@{}l@{}}0\\ (\$0.147)\end{tabular}} &
  {\begin{tabular}[c]{@{}l@{}}0\\ (\$0.091)\end{tabular}} &
  {\begin{tabular}[c]{@{}l@{}}0\\ (\$0)*\\ (\$0.006)\end{tabular}} &
  {\begin{tabular}[c]{@{}l@{}}0\\ (\$0)*\\ (\$0.038)\end{tabular}} &
  {\begin{tabular}[c]{@{}l@{}}0\\ (\$0)*\\ (\$0.022)\end{tabular}} &
  {\begin{tabular}[c]{@{}l@{}}0\\ (\$0)*\\ (\$0.023)\end{tabular}} &
  {\begin{tabular}[c]{@{}l@{}}0\\ (\$0.014)*\\ (\$0.060)\end{tabular}} &
  {\begin{tabular}[c]{@{}l@{}}0\\ (\$0.071)*\\ (\$0.124)\end{tabular}} \\
\textbf{\begin{tabular}[c]{@{}l@{}}ogbn-arxiv\end{tabular}} &

  {\begin{tabular}[c]{@{}l@{}}50\\ (\$1.27)\end{tabular}} &
  {\textbf{\begin{tabular}[c]{@{}l@{}}\underline{87.5}\\ \underline{(\$1.112)}\end{tabular}}} &
  {\begin{tabular}[c]{@{}l@{}}40\\ (\$0.125)\end{tabular}} &
  {\begin{tabular}[c]{@{}l@{}}32\\ (\$0.109)\end{tabular}} &
  {\begin{tabular}[c]{@{}l@{}}25\\ (\$0)*\\ (\$0.011)\end{tabular}} &
  {\begin{tabular}[c]{@{}l@{}}25\\ (\$0)*\\ (\$0.016)\end{tabular}} &
  {\begin{tabular}[c]{@{}l@{}}50\\ (\$0)*\\ (\$0.020)\end{tabular}} &
  {\begin{tabular}[c]{@{}l@{}}50\\ (\$0)*\\ (\$0.015)\end{tabular}} &
  {\begin{tabular}[c]{@{}l@{}}50\\ (\$0.033)*\\ (\$0.067)\end{tabular}} &
  {\begin{tabular}[c]{@{}l@{}}75\\ (\$0.026)*\\ (\$0.045)\end{tabular}} \\
\textbf{\begin{tabular}[c]{@{}l@{}}house-price\end{tabular}} &

  {\begin{tabular}[c]{@{}l@{}}25\\ (\$1.6)\end{tabular}} &
  {\begin{tabular}[c]{@{}l@{}}12.5\\ (\$0.938)\end{tabular}} &
  {\begin{tabular}[c]{@{}l@{}}64\\ (\$0.158)\end{tabular}} &
  {\begin{tabular}[c]{@{}l@{}}76\\ (\$0.093)\end{tabular}} &
  {\begin{tabular}[c]{@{}l@{}}25\\ (\$0)*\\ (\$0.042)\end{tabular}} &
  {\begin{tabular}[c]{@{}l@{}}37.5\\ (\$0)*\\ (\$0.070)\end{tabular}} &
  {\begin{tabular}[c]{@{}l@{}}62.5\\ (\$0)*\\ (\$0.046)\end{tabular}} &
  {\begin{tabular}[c]{@{}l@{}}62.5\\ (\$0)*\\ (\$0.049)\end{tabular}} &
  {\textbf{\begin{tabular}[c]{@{}l@{}}\underline{87.5}\\ \underline{(\$0.091)*}\\ \underline{(\$0.161)}\end{tabular}}} &
  {\begin{tabular}[c]{@{}l@{}}75\\ (\$0.038)*\\ (\$0.153)\end{tabular}} \\
\textbf{\begin{tabular}[c]{@{}l@{}}spaceship-\\ titanic\end{tabular}} &

  {\begin{tabular}[c]{@{}l@{}}25\\ (\$1.42)\end{tabular}} &
  {\begin{tabular}[c]{@{}l@{}}12.5\\ (\$0.85)\end{tabular}} &
  {\begin{tabular}[c]{@{}l@{}}4\\ (\$0.141)\end{tabular}} &
  {\begin{tabular}[c]{@{}l@{}}16\\ (\$0.088)\end{tabular}} &
  {\begin{tabular}[c]{@{}l@{}}37.5\\ (\$0)*\\ (\$0.039)\end{tabular}} &
  {\begin{tabular}[c]{@{}l@{}}37.5\\ (\$0)*\\ (\$0.052)\end{tabular}} &
  {\begin{tabular}[c]{@{}l@{}}75\\ (\$0)*\\ (\$0.077)\end{tabular}} &
  {\textbf{\begin{tabular}[c]{@{}l@{}}\underline{100}\\ \underline{(\$0.0004)*}\\ \underline{(\$0.038)}\end{tabular}}} &
  {\begin{tabular}[c]{@{}l@{}}75\\ (\$0.021)*\\ (\$0.096)\end{tabular}} &
  {\textbf{\begin{tabular}[c]{@{}l@{}}\underline{100}\\ \underline{(\$0.091)*}\\ \underline{(\$0.213)}\end{tabular}}} \\
\textbf{\begin{tabular}[c]{@{}l@{}}parkinsons-\\ disease\end{tabular}} &

  {\textbf{\begin{tabular}[c]{@{}l@{}}\underline{12.5}\\ \underline{(\$2.9)}\end{tabular}}} &
  {\begin{tabular}[c]{@{}l@{}}0\\ (\$1.57)\end{tabular}} &
  {\begin{tabular}[c]{@{}l@{}}0\\ (\$0.287)\end{tabular}} &
  {\begin{tabular}[c]{@{}l@{}}0\\ (\$0.155)\end{tabular}} &
  {\begin{tabular}[c]{@{}l@{}}0\\ (\$0)*\\ (\$0.078)\end{tabular}} &
  {\begin{tabular}[c]{@{}l@{}}0\\ (\$0)*\\ (\$0.065)\end{tabular}} &
  {\begin{tabular}[c]{@{}l@{}}0\\ (\$0)*\\ (\$0.024)\end{tabular}} &
  {\begin{tabular}[c]{@{}l@{}}0\\ (\$0)*\\ (\$0.028)\end{tabular}} &
  {\begin{tabular}[c]{@{}l@{}}0\\ (\$0.099)*\\ (\$0.186)\end{tabular}} &
  {\begin{tabular}[c]{@{}l@{}}0\\ (\$0.107)*\\ (\$0.183)\end{tabular}} \\
\textbf{\begin{tabular}[c]{@{}l@{}}feedback\end{tabular}} &

  {\textbf{\begin{tabular}[c]{@{}l@{}}\underline{37.5}\\ \underline{(\$1.25)}\end{tabular}}} &
  {\begin{tabular}[c]{@{}l@{}}12.5\\ (\$1.15)\end{tabular}} &
  {\begin{tabular}[c]{@{}l@{}}0\\ (\$0.124)\end{tabular}} &
  {\begin{tabular}[c]{@{}l@{}}0\\ (\$0.114)\end{tabular}} &
  {\begin{tabular}[c]{@{}l@{}}0\\ (\$0)*\\ (\$0.058)\end{tabular}} &
  {\begin{tabular}[c]{@{}l@{}}0\\ (\$0)*\\ (\$0.046)\end{tabular}} &
  {\begin{tabular}[c]{@{}l@{}}0\\ (\$0.0005)*\\ (\$0.038)\end{tabular}} &
  {\begin{tabular}[c]{@{}l@{}}0\\ (\$0)*\\ (\$0.030)\end{tabular}} &
  {\begin{tabular}[c]{@{}l@{}}0\\ (\$0.047)*\\ (\$0.110)\end{tabular}} &
  {\begin{tabular}[c]{@{}l@{}}0\\ (\$0.022)*\\ (\$0.068)\end{tabular}} \\
\textbf{\begin{tabular}[c]{@{}l@{}}llama-\\ inference\end{tabular}} &

  {\begin{tabular}[c]{@{}l@{}}0\\ (\$1.46)\end{tabular}} &
  {\begin{tabular}[c]{@{}l@{}}0\\ (\$0.927)\end{tabular}} &
  {\begin{tabular}[c]{@{}l@{}}0\\ (\$0.145)\end{tabular}} &
  {\begin{tabular}[c]{@{}l@{}}0\\ (\$0.092)\end{tabular}} &
  {\begin{tabular}[c]{@{}l@{}}0\\ (\$0)*\\ (\$0.030)\end{tabular}} &
  {\begin{tabular}[c]{@{}l@{}}0\\ (\$0)*\\ (\$0.059)\end{tabular}} &
  {\begin{tabular}[c]{@{}l@{}}0\\ (\$0.0003)*\\ (\$0.032)\end{tabular}} &
  {\begin{tabular}[c]{@{}l@{}}0\\ (\$0)*\\ (\$0.034)\end{tabular}} &
  {\begin{tabular}[c]{@{}l@{}}0\\ (\$0.027)*\\ (\$0.074)\end{tabular}} &
  {\begin{tabular}[c]{@{}l@{}}0\\ (\$0.055)*\\ (\$0.093)\end{tabular}} \\
\textbf{\begin{tabular}[c]{@{}l@{}}vectorization\end{tabular}} &

  {\begin{tabular}[c]{@{}l@{}}0\\ (\$1.16)\end{tabular}} &
  {\begin{tabular}[c]{@{}l@{}}0\\ (\$1.23)\end{tabular}} &
  {\begin{tabular}[c]{@{}l@{}}0\\ (\$0.114)\end{tabular}} &
  {\begin{tabular}[c]{@{}l@{}}0\\ (\$0.121)\end{tabular}} &
  {\begin{tabular}[c]{@{}l@{}}0\\ (\$0)*\\ (\$0.062)\end{tabular}} &
  {\begin{tabular}[c]{@{}l@{}}12.5\\ (\$0)*\\ (\$0.058)\end{tabular}} &
  {\begin{tabular}[c]{@{}l@{}}0\\ (\$0)*\\ (\$0.039)\end{tabular}} &
  {\begin{tabular}[c]{@{}l@{}}12.5\\ (\$0)*\\ (\$0.025)\end{tabular}} &
  {\begin{tabular}[c]{@{}l@{}}0\\ (\$0.017)*\\ (\$0.077)\end{tabular}} &
  {\textbf{\begin{tabular}[c]{@{}l@{}}\underline{75}\\ \underline{(\$0.004)*}\\ \underline{(\$0.092)}\end{tabular}}} \\
\textbf{\begin{tabular}[c]{@{}l@{}}CLRS\end{tabular}} &

  {\begin{tabular}[c]{@{}l@{}}12.5\\ (\$0.523)\end{tabular}} &
  {\begin{tabular}[c]{@{}l@{}}50\\ (\$0.43)\end{tabular}} &
  {\textbf{\begin{tabular}[c]{@{}l@{}}\underline{52}\\ \underline{(\$0.052)}\end{tabular}}} &
  {\begin{tabular}[c]{@{}l@{}}40\\ (\$0.042)\end{tabular}} &
  {\begin{tabular}[c]{@{}l@{}}0\\ (\$0)*\\ (\$0.006)\end{tabular}} &
  {\begin{tabular}[c]{@{}l@{}}0\\ (\$0)*\\ (\$0.006*)\end{tabular}} &
  {\begin{tabular}[c]{@{}l@{}}0\\ (\$0.0003)*\\ (\$0.033)\end{tabular}} &
  {\begin{tabular}[c]{@{}l@{}}12.5\\ (\$0)*\\ (\$0.022)\end{tabular}} &
  {\begin{tabular}[c]{@{}l@{}}0\\ (\$0.027)*\\ (\$0.050)\end{tabular}} &
  {\begin{tabular}[c]{@{}l@{}}0\\ (\$0.046)*\\ (\$0.138)\end{tabular}} \\
\textbf{\begin{tabular}[c]{@{}l@{}}babylm\end{tabular}} &

  {\begin{tabular}[c]{@{}l@{}}0\\ (\$0.818)\end{tabular}} &
  {\begin{tabular}[c]{@{}l@{}}0\\ (\$0.637)\end{tabular}} &
  {\begin{tabular}[c]{@{}l@{}}0\\ (\$0.081)\end{tabular}} &
  {\begin{tabular}[c]{@{}l@{}}0\\ (\$0.063)\end{tabular}} &
  {\begin{tabular}[c]{@{}l@{}}0\\ (\$0)*\\ (\$0.032)\end{tabular}} &
  {\begin{tabular}[c]{@{}l@{}}0\\ (\$0)*\\ (\$0*.027)\end{tabular}} &
  {\begin{tabular}[c]{@{}l@{}}0\\ (\$0.0025)*\\ (\$0.046)\end{tabular}} &
  {\begin{tabular}[c]{@{}l@{}}0\\ (\$0.001)*\\ (\$0.024)\end{tabular}} &
  {\begin{tabular}[c]{@{}l@{}}0\\ (\$0.086)*\\ (\$0.147)\end{tabular}} &
  {\begin{tabular}[c]{@{}l@{}}0\\ (\$0.078)*\\ (\$0.113)\end{tabular}} \\ \midrule
\textbf{Average} &

  {\begin{tabular}[c]{@{}l@{}}18.18\\ (\$1.315)\end{tabular}} &
  {\begin{tabular}[c]{@{}l@{}}22.72\\ (\$0.931)\end{tabular}} &
  {\begin{tabular}[c]{@{}l@{}}15.27\\ (\$0.13)\end{tabular}} &
  {\begin{tabular}[c]{@{}l@{}}20\\ (\$0.092)\end{tabular}} &
  {\begin{tabular}[c]{@{}l@{}}9.09\\ (\$0)*\\ (\$0.034)\end{tabular}} &
  {\begin{tabular}[c]{@{}l@{}}10.23\\ (\$0)*\\ (\$0.044)\end{tabular}} &
  {\begin{tabular}[c]{@{}l@{}}22.72\\ (\$0.0003)*\\ (\$0.036)\end{tabular}} &
  {\begin{tabular}[c]{@{}l@{}}22.72\\ (\$0.0001)*\\ (\$0.032)\end{tabular}} &
  {\begin{tabular}[c]{@{}l@{}}26.14\\ (\$0.047)*\\ (\$0.102)\end{tabular}} &
  {\textbf{\begin{tabular}[c]{@{}l@{}}\underline{32.95}\\ \underline{(\$0.054)*}\\ \underline{(\$0.120)}\end{tabular}}} \\
  \bottomrule
\end{tabular}%
}
\caption{Each cell depicts Success Rate in \%: \% successful runs,  (\$):  Average cost per run, * - considering free Gemini Pro API calls; G - GPT4 , C - Claude V1.0, Ge - Gemini Pro, Co - CodeLlama Instruct 34b, Ch - ChatGPT (GPT 3.5 Turbo), R -  Retrieval: The agent can retrieve and summarize relevant information from long-term memory in research logs. In no-retrieval setting this functionality is disabled.}
\label{tab:main-results}
\end{table*}

\subsection{Results and Discussion}

We address key Research Questions (RQ) based on the results in table \ref{tab:main-results}.

\subsubsection{RQ1 - Do no-cost and low-cost LLM single-agents sacrifice performance for cost savings?}

We observe a significant drop in the performance when we use a purely no-cost or low-cost LLM (Ge, Co, and Mx) in single-agent setting across all tasks as opposed to LLM single agents using GPT4 or Claude presented in \citet{huang2023benchmarking}. We observe CodeLlama (Co) and Mixtral (Mx) are unable to produce any successful runs leading to average 0\% success rate. Thus, we omit them from the results in table \ref{tab:main-results}. We see that Mixtral is almost never able to adhere to the required response format mentioned in section \ref{sec:multi-llm} across all runs which leads to termination due to maximum retry limit being exceeded. However, we observe that Gemini-Pro (Ge + R) is able to produce  successful runs in single agent setting yielding non-zero performance for some tasks (cifar10, ogbn-arxiv, house-price and spaceship-titanic) but zero for others.  Visual inspection reveals that there is not much difference between the quality of responses in internal action-related LLM calls between Gemini-Pro and CodeLllama, with the former being slightly superior. However, we observe that Code Llama is unable to produce any successful runs due to its inability to plan effectively.

\subsubsection{RQ2 - How do profiling and cascades affect performance and cost?}

From table \ref{tab:main-results}, it can be seen that profiling with Gemini-Pro as base LLM and ChatGPT (GPT-3.5-turbo) in cascade (Ge + Ch) both with and without retriever setting, significantly increases success rate for many tasks, namely cifar10, ogbn-arxiv, house-price, spaceship-titanic and vectorization when compared with Ge as single agent. However, for imdb, parkinsons-disease, feedback, llama-inference, CLRS and babylm tasks, the performance still remains zero due to their complex nature. Overall average success rate of profiling and cascade is comparable with GPT4 (G) performance. These improvements are obtained at almost 100\% cost reduction for no-cost Gemini-Pro (From \$1.315 for G + R and \$0.13 for C + R to \$0.0003 for Ge + Ch + R and from \$0.931 for G and \$0.092 for C to \$0.0001 for Ge + Ch). This is because inference for GPT-3.5-turbo is much cheaper leading to minuscule costs. However, qualitative analysis shows that GPT-3.5-turbo often fails to adhere to the required response format mentioned in section \ref{sec:multi-llm} leading to further retries. Thus, we shift to GPT4 for cascades for subsequent runs.

\subsubsection{RQ3 - How does adding ask-the-expert lifelines to profiling and cascade affect performance and cost?}

Table \ref{tab:main-results} shows that access to GPT4 ask-the-expert lifeline calls as part of our proposed system (Ge + G  and Ge + G  + R) improves success rate for cifar10, ogbn-arxiv, house-price, spaceship-titanic and vectorization tasks when compared with Ge + Ch+ R and Ge + Ch. Additionally, we observe improvements in success rate when compared with G + R and G for cifar10, house-price, spaceship-titanic and vectorization tasks at a cost reduction ranging from 90-99\% across tasks for no-cost Gemini-Pro. On an average the Profiling + Cascade + Expert setting and using GPT4 for cascade and expert gives 43.78\% and 71.19\% improvement in retrieval setting and 45.02\% and 64.75\% improvement in non-retrieval setting over GPT4 and Claude as single agent, respectively. This comes at 96.43\%  and 63.85\% cost reduction for no-cost Gemini-Pro in retrieval setting and 94.2\%  and 41.3\% reduction in cost in non-retrieval setting over GPT4 and Claude, respectively, demonstrating the efficacy of our approach. 
Qualitative analysis reveals that there are instances when the planner can successfully identify that it is stuck, and calls the expert to get itself 'unstuck'. For example, if a particular edit does not lead to proper execution after multiple steps, and if the expert is called, it takes a different action such as understanding some part of the file better before making further edits.


\subsubsection{RQ4 - How does retrieval from logs affect performance and cost?}

In accordance with the findings of \citet{huang2023benchmarking}, our observations indicate that while access to retrieval from logs proves effective for certain tasks, for others, disabling it yields better results. One interesting case to note is of cifar10.  GPT4 (G) has a greater success rate than G + R. 
\citet{huang2023benchmarking} justify this by stating that since cifar10 is a comparitively easier task and the long-term memory context become a distraction. But the trend gets reversed in the case of Gemini-Pro with GPT4 in cascade and expert setting.  Ge + G + R has a greater success rate than Ge + G. 
This can be due to the differences in pretraining and actual data seen by Gemini-Pro and GPT4. Similar to success rate, cost is greater with retrieval for some tasks, whereas it is greater without retrieval for others. Average cost per run for Ge + G (\$0.054) is greater than that for Ge + G + R (\$0.047) for no-cost Gemini-Pro.
\section{Related Work}

\subsection{LLM Agents for nuanced code-related tasks}

With the rising capabilities of Large Language Models, there have been many advances in employing LLM Agents for code-related tasks. Initial works relied primarily on HumanEval \citep{chen2021evaluating} and MBXP \citep{athiwaratkun2023multilingual} for benchmarking their approaches. 
However, recent works have started moving towards more nuanced and real-world benchmarks.
\citet{wang2024executable} present CodeAct - a Python code database containing tasks related to API handling, library usage, etc. They also present CodeActAgent that is finetuned on Llama-2 \citep{touvron2023llama} and Mistral-7b \citep{jiang2023mistral} using CodeActInstruct, an instruction tuning dataset consisting of multi-turn interactions.
AgentBench \citep{liu2023agentbench} also introduces three code agent environments based on operating system, database and knowledge graph related problems for evaluating LLM agents on these tasks.
There has also been some amount of work done for tackling more complex problems. \citet{zhang2024codeagent} construct CodeAgentBench, a dataset containing real-world repo-level coding challenges. They also propose CodeAgent to tackle the tasks in this dataset utilizing external information retrieval, code implementation and testing tools at its disposal.

\citet{huang2023benchmarking} construct MLAgentBench, explained in Section \ref{sec:mlagentbench}, that consists of machine learning tasks. They also propose a single agent based system to solve these tasks using various tools such as code execution, code editing, etc. 
\citet{tang2024mlbenchevaluatinglargelanguage} explores the use of LLM Agents with repository-level code. However, their approach does not involve direct code generation. Instead, they focus on creating shell scripts to utilize pre-existing code files with appropriate arguments. On a similar note, CodePlan \citep{bairi2023codeplanrepositorylevelcodingusing} involves framing repository-level coding as a planning problem in the form of a chain of edits. Additionally, NL2Repo \citep{zan2024codesnaturallanguagecode} presents a system for generating codebases from the ground-up, using natural language specifications.

\subsection{LLM Multi-Agents}

While LLM single-agents have demonstrated remarkable capabilities in various tasks, the complexity of real-world challenges often necessitates collaborative approaches. In this context, the exploration of LLM multi-agent systems has emerged as a promising avenue to enhance problem-solving capabilities and address intricate tasks more effectively.
ChatDev \citep{qian2023communicative} and MetaGPT \citep{hong2023metagpt} are recent multi-agent systems introduced for tackling software development tasks. 
AutoGen \citep{wu2023autogen} is a more generic framework for developing LLM multi-agents and has been evaluated on multiple tasks ranging from math problem solving to retrieval augmented chatting
\citet{li2024Agents, shen2024small} also present frameworks for using multiple LLM agents in a collaborative manner for tasks such as reasoning and tool-learning.
\citet{wang2023fusing, ding2024hybrid} propose combining multiple LLM expert agents to solve a variety of tasks including text summarization and question answering.
To save up on costs, \citet{chen2023frugalgpt, yue2024large, zhang2023ecoassistant} propose using cascading LLMs in increasing order of cost and capability for diverse tasks such as reasoning, query answering and news prediction. Our focus is on providing cost-effective solution for ML engineering tasks using multi-agent system with no-cost open-source LLM as the base with paid LLM with higher expertise in cascade.
\section{Conclusion}

In this work, we propose BudgetMLAgent - an LLM Multi-Agent system for solving machine learning tasks in a cost-effective manner without hampering the system performance. Our primary investigations on tasks defined in MLAgentBench dataset, with Single-Agent systems with purely no-cost models give zero success rates for CodeLlama and Mixtral and very poor success rates with Gemini-Pro (9.09\% with access to the complete logs and 10.23\% with short-term access), as compared to paid models such as GPT4 and ClaudeV1.0 (18.18\% and 22.72\%, 15.27\% and 20\% respectively). We subsequently propose a multi-agent framework, BudgetMLAgent, using no-cost Gemini-Pro as the base LLM, leveraging (i) profiling for a planner and multiple worker agents interacting with the ML code generation environment using distinct actions, (ii) cascades to LLMs with more expertise such as GPT-3.5-turbo and GPT4 and (iii) our novel ask-the-expert GPT4 lifelines. BudgetMLAgent results in improving the success rates for MLAgentBench tasks (26.14\% and 32.95\% respectively) along with significant cost reductions when compared with GPT4 and ClaudeV1.0 based Single-Agent systems (96.43\% and 63.85\% with access to the complete logs, 94.2\% and 41.3\% reduction with short-term access respectively).



\begin{thebibliography}{50}


\ifx \showCODEN    \undefined \def \showCODEN     #1{\unskip}     \fi
\ifx \showDOI      \undefined \def \showDOI       #1{#1}\fi
\ifx \showISBNx    \undefined \def \showISBNx     #1{\unskip}     \fi
\ifx \showISBNxiii \undefined \def \showISBNxiii  #1{\unskip}     \fi
\ifx \showISSN     \undefined \def \showISSN      #1{\unskip}     \fi
\ifx \showLCCN     \undefined \def \showLCCN      #1{\unskip}     \fi
\ifx \shownote     \undefined \def \shownote      #1{#1}          \fi
\ifx \showarticletitle \undefined \def \showarticletitle #1{#1}   \fi
\ifx \showURL      \undefined \def \showURL       {\relax}        \fi
\providecommand\bibfield[2]{#2}
\providecommand\bibinfo[2]{#2}
\providecommand\natexlab[1]{#1}
\providecommand\showeprint[2][]{arXiv:#2}

\bibitem[Addison~Howard(2022)]%
        {spaceship-titanic}
\bibfield{author}{\bibinfo{person}{Ryan~Holbrook Addison~Howard, Ashley~Chow}.} \bibinfo{year}{2022}\natexlab{}.
\newblock \bibinfo{title}{Spaceship Titanic}.
\newblock
\newblock
\urldef\tempurl%
\url{https://kaggle.com/competitions/spaceship-titanic}
\showURL{%
\tempurl}


\bibitem[Anna~Montoya(2016)]%
        {house-prices-advanced-regression-techniques}
\bibfield{author}{\bibinfo{person}{DataCanary Anna~Montoya}.} \bibinfo{year}{2016}\natexlab{}.
\newblock \bibinfo{title}{House Prices - Advanced Regression Techniques}.
\newblock
\newblock
\urldef\tempurl%
\url{https://kaggle.com/competitions/house-prices-advanced-regression-techniques}
\showURL{%
\tempurl}


\bibitem[Athiwaratkun et~al\mbox{.}(2023)]%
        {athiwaratkun2023multilingual}
\bibfield{author}{\bibinfo{person}{Ben Athiwaratkun}, \bibinfo{person}{Sanjay~Krishna Gouda}, \bibinfo{person}{Zijian Wang}, \bibinfo{person}{Xiaopeng Li}, \bibinfo{person}{Yuchen Tian}, \bibinfo{person}{Ming Tan}, \bibinfo{person}{Wasi~Uddin Ahmad}, \bibinfo{person}{Shiqi Wang}, \bibinfo{person}{Qing Sun}, \bibinfo{person}{Mingyue Shang}, \bibinfo{person}{Sujan~Kumar Gonugondla}, \bibinfo{person}{Hantian Ding}, \bibinfo{person}{Varun Kumar}, \bibinfo{person}{Nathan Fulton}, \bibinfo{person}{Arash Farahani}, \bibinfo{person}{Siddhartha Jain}, \bibinfo{person}{Robert Giaquinto}, \bibinfo{person}{Haifeng Qian}, \bibinfo{person}{Murali~Krishna Ramanathan}, \bibinfo{person}{Ramesh Nallapati}, \bibinfo{person}{Baishakhi Ray}, \bibinfo{person}{Parminder Bhatia}, \bibinfo{person}{Sudipta Sengupta}, \bibinfo{person}{Dan Roth}, {and} \bibinfo{person}{Bing Xiang}.} \bibinfo{year}{2023}\natexlab{}.
\newblock \bibinfo{title}{Multi-lingual Evaluation of Code Generation Models}.
\newblock
\newblock
\showeprint[arxiv]{2210.14868}~[cs.LG]


\bibitem[Bairi et~al\mbox{.}(2023)]%
        {bairi2023codeplanrepositorylevelcodingusing}
\bibfield{author}{\bibinfo{person}{Ramakrishna Bairi}, \bibinfo{person}{Atharv Sonwane}, \bibinfo{person}{Aditya Kanade}, \bibinfo{person}{Vageesh~D C}, \bibinfo{person}{Arun Iyer}, \bibinfo{person}{Suresh Parthasarathy}, \bibinfo{person}{Sriram Rajamani}, \bibinfo{person}{B. Ashok}, {and} \bibinfo{person}{Shashank Shet}.} \bibinfo{year}{2023}\natexlab{}.
\newblock \bibinfo{title}{CodePlan: Repository-level Coding using LLMs and Planning}.
\newblock
\newblock
\showeprint[arxiv]{2309.12499}~[cs.SE]
\urldef\tempurl%
\url{https://arxiv.org/abs/2309.12499}
\showURL{%
\tempurl}


\bibitem[Chen et~al\mbox{.}(2023)]%
        {chen2023frugalgpt}
\bibfield{author}{\bibinfo{person}{Lingjiao Chen}, \bibinfo{person}{Matei Zaharia}, {and} \bibinfo{person}{James Zou}.} \bibinfo{year}{2023}\natexlab{}.
\newblock \bibinfo{title}{FrugalGPT: How to Use Large Language Models While Reducing Cost and Improving Performance}.
\newblock
\newblock
\showeprint[arxiv]{2305.05176}~[cs.LG]


\bibitem[Chen et~al\mbox{.}(2021)]%
        {chen2021evaluating}
\bibfield{author}{\bibinfo{person}{Mark Chen}, \bibinfo{person}{Jerry Tworek}, \bibinfo{person}{Heewoo Jun}, \bibinfo{person}{Qiming Yuan}, \bibinfo{person}{Henrique~Ponde de Oliveira~Pinto}, \bibinfo{person}{Jared Kaplan}, \bibinfo{person}{Harri Edwards}, \bibinfo{person}{Yuri Burda}, \bibinfo{person}{Nicholas Joseph}, \bibinfo{person}{Greg Brockman}, \bibinfo{person}{Alex Ray}, \bibinfo{person}{Raul Puri}, \bibinfo{person}{Gretchen Krueger}, \bibinfo{person}{Michael Petrov}, \bibinfo{person}{Heidy Khlaaf}, \bibinfo{person}{Girish Sastry}, \bibinfo{person}{Pamela Mishkin}, \bibinfo{person}{Brooke Chan}, \bibinfo{person}{Scott Gray}, \bibinfo{person}{Nick Ryder}, \bibinfo{person}{Mikhail Pavlov}, \bibinfo{person}{Alethea Power}, \bibinfo{person}{Lukasz Kaiser}, \bibinfo{person}{Mohammad Bavarian}, \bibinfo{person}{Clemens Winter}, \bibinfo{person}{Philippe Tillet}, \bibinfo{person}{Felipe~Petroski Such}, \bibinfo{person}{Dave Cummings}, \bibinfo{person}{Matthias Plappert}, \bibinfo{person}{Fotios Chantzis},
  \bibinfo{person}{Elizabeth Barnes}, \bibinfo{person}{Ariel Herbert-Voss}, \bibinfo{person}{William~Hebgen Guss}, \bibinfo{person}{Alex Nichol}, \bibinfo{person}{Alex Paino}, \bibinfo{person}{Nikolas Tezak}, \bibinfo{person}{Jie Tang}, \bibinfo{person}{Igor Babuschkin}, \bibinfo{person}{Suchir Balaji}, \bibinfo{person}{Shantanu Jain}, \bibinfo{person}{William Saunders}, \bibinfo{person}{Christopher Hesse}, \bibinfo{person}{Andrew~N. Carr}, \bibinfo{person}{Jan Leike}, \bibinfo{person}{Josh Achiam}, \bibinfo{person}{Vedant Misra}, \bibinfo{person}{Evan Morikawa}, \bibinfo{person}{Alec Radford}, \bibinfo{person}{Matthew Knight}, \bibinfo{person}{Miles Brundage}, \bibinfo{person}{Mira Murati}, \bibinfo{person}{Katie Mayer}, \bibinfo{person}{Peter Welinder}, \bibinfo{person}{Bob McGrew}, \bibinfo{person}{Dario Amodei}, \bibinfo{person}{Sam McCandlish}, \bibinfo{person}{Ilya Sutskever}, {and} \bibinfo{person}{Wojciech Zaremba}.} \bibinfo{year}{2021}\natexlab{}.
\newblock \bibinfo{title}{Evaluating Large Language Models Trained on Code}.
\newblock
\newblock
\showeprint[arxiv]{2107.03374}~[cs.LG]


\bibitem[Ding et~al\mbox{.}(2024)]%
        {ding2024hybrid}
\bibfield{author}{\bibinfo{person}{Dujian Ding}, \bibinfo{person}{Ankur Mallick}, \bibinfo{person}{Chi Wang}, \bibinfo{person}{Robert Sim}, \bibinfo{person}{Subhabrata Mukherjee}, \bibinfo{person}{Victor R{\"u}hle}, \bibinfo{person}{Laks V.~S. Lakshmanan}, {and} \bibinfo{person}{Ahmed~Hassan Awadallah}.} \bibinfo{year}{2024}\natexlab{}.
\newblock \showarticletitle{Hybrid {LLM}: Cost-Efficient and Quality-Aware Query Routing}. In \bibinfo{booktitle}{\emph{The Twelfth International Conference on Learning Representations}}.
\newblock
\urldef\tempurl%
\url{https://openreview.net/forum?id=02f3mUtqnM}
\showURL{%
\tempurl}


\bibitem[Fang et~al\mbox{.}(2024)]%
        {fang2024large}
\bibfield{author}{\bibinfo{person}{Xi Fang}, \bibinfo{person}{Weijie Xu}, \bibinfo{person}{Fiona~Anting Tan}, \bibinfo{person}{Jiani Zhang}, \bibinfo{person}{Ziqing Hu}, \bibinfo{person}{Yanjun Qi}, \bibinfo{person}{Scott Nickleach}, \bibinfo{person}{Diego Socolinsky}, \bibinfo{person}{Srinivasan Sengamedu}, {and} \bibinfo{person}{Christos Faloutsos}.} \bibinfo{year}{2024}\natexlab{}.
\newblock \bibinfo{title}{Large Language Models(LLMs) on Tabular Data: Prediction, Generation, and Understanding -- A Survey}.
\newblock
\newblock
\showeprint[arxiv]{2402.17944}~[cs.CL]


\bibitem[Franklin et~al\mbox{.}(2022)]%
        {feedback-prize-english-language-learning}
\bibfield{author}{\bibinfo{person}{Alex Franklin}, \bibinfo{person}{Maggie}, \bibinfo{person}{Meg Benner}, \bibinfo{person}{Natalie Rambis}, \bibinfo{person}{Perpetual Baffour}, \bibinfo{person}{Ryan Holbrook}, \bibinfo{person}{Scott Crossley}, {and} \bibinfo{person}{Ulrich Boser}.} \bibinfo{year}{2022}\natexlab{}.
\newblock \bibinfo{title}{Feedback Prize - English Language Learning}.
\newblock
\newblock
\urldef\tempurl%
\url{https://kaggle.com/competitions/feedback-prize-english-language-learning}
\showURL{%
\tempurl}


\bibitem[Guo et~al\mbox{.}(2024)]%
        {guo2024deepseekcoder}
\bibfield{author}{\bibinfo{person}{Daya Guo}, \bibinfo{person}{Qihao Zhu}, \bibinfo{person}{Dejian Yang}, \bibinfo{person}{Zhenda Xie}, \bibinfo{person}{Kai Dong}, \bibinfo{person}{Wentao Zhang}, \bibinfo{person}{Guanting Chen}, \bibinfo{person}{Xiao Bi}, \bibinfo{person}{Y. Wu}, \bibinfo{person}{Y.~K. Li}, \bibinfo{person}{Fuli Luo}, \bibinfo{person}{Yingfei Xiong}, {and} \bibinfo{person}{Wenfeng Liang}.} \bibinfo{year}{2024}\natexlab{}.
\newblock \bibinfo{title}{DeepSeek-Coder: When the Large Language Model Meets Programming -- The Rise of Code Intelligence}.
\newblock
\newblock
\showeprint[arxiv]{2401.14196}~[cs.SE]


\bibitem[He et~al\mbox{.}(2021)]%
        {He_2021}
\bibfield{author}{\bibinfo{person}{Xin He}, \bibinfo{person}{Kaiyong Zhao}, {and} \bibinfo{person}{Xiaowen Chu}.} \bibinfo{year}{2021}\natexlab{}.
\newblock \showarticletitle{AutoML: A survey of the state-of-the-art}.
\newblock \bibinfo{journal}{\emph{Knowledge-Based Systems}}  \bibinfo{volume}{212} (\bibinfo{date}{Jan.} \bibinfo{year}{2021}), \bibinfo{pages}{106622}.
\newblock
\showISSN{0950-7051}
\urldef\tempurl%
\url{https://doi.org/10.1016/j.knosys.2020.106622}
\showDOI{\tempurl}


\bibitem[Hong et~al\mbox{.}(2023)]%
        {hong2023metagpt}
\bibfield{author}{\bibinfo{person}{Sirui Hong}, \bibinfo{person}{Mingchen Zhuge}, \bibinfo{person}{Jonathan Chen}, \bibinfo{person}{Xiawu Zheng}, \bibinfo{person}{Yuheng Cheng}, \bibinfo{person}{Ceyao Zhang}, \bibinfo{person}{Jinlin Wang}, \bibinfo{person}{Zili Wang}, \bibinfo{person}{Steven Ka~Shing Yau}, \bibinfo{person}{Zijuan Lin}, \bibinfo{person}{Liyang Zhou}, \bibinfo{person}{Chenyu Ran}, \bibinfo{person}{Lingfeng Xiao}, \bibinfo{person}{Chenglin Wu}, {and} \bibinfo{person}{Jürgen Schmidhuber}.} \bibinfo{year}{2023}\natexlab{}.
\newblock \bibinfo{title}{MetaGPT: Meta Programming for A Multi-Agent Collaborative Framework}.
\newblock
\newblock
\showeprint[arxiv]{2308.00352}~[cs.AI]


\bibitem[Hu et~al\mbox{.}(2021)]%
        {hu2021open}
\bibfield{author}{\bibinfo{person}{Weihua Hu}, \bibinfo{person}{Matthias Fey}, \bibinfo{person}{Marinka Zitnik}, \bibinfo{person}{Yuxiao Dong}, \bibinfo{person}{Hongyu Ren}, \bibinfo{person}{Bowen Liu}, \bibinfo{person}{Michele Catasta}, {and} \bibinfo{person}{Jure Leskovec}.} \bibinfo{year}{2021}\natexlab{}.
\newblock \bibinfo{title}{Open Graph Benchmark: Datasets for Machine Learning on Graphs}.
\newblock
\newblock
\showeprint[arxiv]{2005.00687}~[cs.LG]


\bibitem[Huang et~al\mbox{.}(2024)]%
        {huang2024agentcoder}
\bibfield{author}{\bibinfo{person}{Dong Huang}, \bibinfo{person}{Qingwen Bu}, \bibinfo{person}{Jie~M. Zhang}, \bibinfo{person}{Michael Luck}, {and} \bibinfo{person}{Heming Cui}.} \bibinfo{year}{2024}\natexlab{}.
\newblock \bibinfo{title}{AgentCoder: Multi-Agent-based Code Generation with Iterative Testing and Optimisation}.
\newblock
\newblock
\showeprint[arxiv]{2312.13010}~[cs.CL]


\bibitem[Huang and Chang(2023)]%
        {huang2023reasoning}
\bibfield{author}{\bibinfo{person}{Jie Huang} {and} \bibinfo{person}{Kevin Chen-Chuan Chang}.} \bibinfo{year}{2023}\natexlab{}.
\newblock \bibinfo{title}{Towards Reasoning in Large Language Models: A Survey}.
\newblock
\newblock
\showeprint[arxiv]{2212.10403}~[cs.CL]


\bibitem[Huang et~al\mbox{.}(2023)]%
        {huang2023benchmarking}
\bibfield{author}{\bibinfo{person}{Qian Huang}, \bibinfo{person}{Jian Vora}, \bibinfo{person}{Percy Liang}, {and} \bibinfo{person}{Jure Leskovec}.} \bibinfo{year}{2023}\natexlab{}.
\newblock \bibinfo{title}{Benchmarking Large Language Models As AI Research Agents}.
\newblock
\newblock
\showeprint[arxiv]{2310.03302}~[cs.LG]


\bibitem[Jiang et~al\mbox{.}(2023)]%
        {jiang2023mistral}
\bibfield{author}{\bibinfo{person}{Albert~Q. Jiang}, \bibinfo{person}{Alexandre Sablayrolles}, \bibinfo{person}{Arthur Mensch}, \bibinfo{person}{Chris Bamford}, \bibinfo{person}{Devendra~Singh Chaplot}, \bibinfo{person}{Diego de~las Casas}, \bibinfo{person}{Florian Bressand}, \bibinfo{person}{Gianna Lengyel}, \bibinfo{person}{Guillaume Lample}, \bibinfo{person}{Lucile Saulnier}, \bibinfo{person}{Lélio~Renard Lavaud}, \bibinfo{person}{Marie-Anne Lachaux}, \bibinfo{person}{Pierre Stock}, \bibinfo{person}{Teven~Le Scao}, \bibinfo{person}{Thibaut Lavril}, \bibinfo{person}{Thomas Wang}, \bibinfo{person}{Timothée Lacroix}, {and} \bibinfo{person}{William~El Sayed}.} \bibinfo{year}{2023}\natexlab{}.
\newblock \bibinfo{title}{Mistral 7B}.
\newblock
\newblock
\showeprint[arxiv]{2310.06825}~[cs.CL]


\bibitem[Jiang et~al\mbox{.}(2024)]%
        {jiang2024mixtral}
\bibfield{author}{\bibinfo{person}{Albert~Q. Jiang}, \bibinfo{person}{Alexandre Sablayrolles}, \bibinfo{person}{Antoine Roux}, \bibinfo{person}{Arthur Mensch}, \bibinfo{person}{Blanche Savary}, \bibinfo{person}{Chris Bamford}, \bibinfo{person}{Devendra~Singh Chaplot}, \bibinfo{person}{Diego de~las Casas}, \bibinfo{person}{Emma~Bou Hanna}, \bibinfo{person}{Florian Bressand}, \bibinfo{person}{Gianna Lengyel}, \bibinfo{person}{Guillaume Bour}, \bibinfo{person}{Guillaume Lample}, \bibinfo{person}{Lélio~Renard Lavaud}, \bibinfo{person}{Lucile Saulnier}, \bibinfo{person}{Marie-Anne Lachaux}, \bibinfo{person}{Pierre Stock}, \bibinfo{person}{Sandeep Subramanian}, \bibinfo{person}{Sophia Yang}, \bibinfo{person}{Szymon Antoniak}, \bibinfo{person}{Teven~Le Scao}, \bibinfo{person}{Théophile Gervet}, \bibinfo{person}{Thibaut Lavril}, \bibinfo{person}{Thomas Wang}, \bibinfo{person}{Timothée Lacroix}, {and} \bibinfo{person}{William~El Sayed}.} \bibinfo{year}{2024}\natexlab{}.
\newblock \bibinfo{title}{Mixtral of Experts}.
\newblock
\newblock
\showeprint[arxiv]{2401.04088}~[cs.LG]


\bibitem[Krizhevsky(2009)]%
        {Krizhevsky2009LearningML}
\bibfield{author}{\bibinfo{person}{Alex Krizhevsky}.} \bibinfo{year}{2009}\natexlab{}.
\newblock \showarticletitle{Learning Multiple Layers of Features from Tiny Images}.
\newblock
\urldef\tempurl%
\url{https://api.semanticscholar.org/CorpusID:18268744}
\showURL{%
\tempurl}


\bibitem[Leslie~Kirsch and Dardov(2023)]%
        {amp-parkinsons-disease-progression-prediction}
\bibfield{author}{\bibinfo{person}{Stacey~Adam Leslie~Kirsch, Sohier~Dane} {and} \bibinfo{person}{Victoria Dardov}.} \bibinfo{year}{2023}\natexlab{}.
\newblock \bibinfo{title}{AMP®-Parkinson's Disease Progression Prediction}.
\newblock
\newblock
\urldef\tempurl%
\url{https://kaggle.com/competitions/amp-parkinsons-disease-progression-prediction}
\showURL{%
\tempurl}


\bibitem[Li et~al\mbox{.}(2024)]%
        {li2024Agents}
\bibfield{author}{\bibinfo{person}{Junyou Li}, \bibinfo{person}{Qin Zhang}, \bibinfo{person}{Yangbin Yu}, \bibinfo{person}{Qiang Fu}, {and} \bibinfo{person}{Deheng Ye}.} \bibinfo{year}{2024}\natexlab{}.
\newblock \bibinfo{title}{More Agents Is All You Need}.
\newblock
\newblock
\showeprint[arxiv]{2402.05120}~[cs.CL]


\bibitem[Liu et~al\mbox{.}(2023)]%
        {liu2023agentbench}
\bibfield{author}{\bibinfo{person}{Xiao Liu}, \bibinfo{person}{Hao Yu}, \bibinfo{person}{Hanchen Zhang}, \bibinfo{person}{Yifan Xu}, \bibinfo{person}{Xuanyu Lei}, \bibinfo{person}{Hanyu Lai}, \bibinfo{person}{Yu Gu}, \bibinfo{person}{Hangliang Ding}, \bibinfo{person}{Kaiwen Men}, \bibinfo{person}{Kejuan Yang}, \bibinfo{person}{Shudan Zhang}, \bibinfo{person}{Xiang Deng}, \bibinfo{person}{Aohan Zeng}, \bibinfo{person}{Zhengxiao Du}, \bibinfo{person}{Chenhui Zhang}, \bibinfo{person}{Sheng Shen}, \bibinfo{person}{Tianjun Zhang}, \bibinfo{person}{Yu Su}, \bibinfo{person}{Huan Sun}, \bibinfo{person}{Minlie Huang}, \bibinfo{person}{Yuxiao Dong}, {and} \bibinfo{person}{Jie Tang}.} \bibinfo{year}{2023}\natexlab{}.
\newblock \bibinfo{title}{AgentBench: Evaluating LLMs as Agents}.
\newblock
\newblock
\showeprint[arxiv]{2308.03688}~[cs.AI]


\bibitem[Maas et~al\mbox{.}(2011)]%
        {maas-etal-2011-learning}
\bibfield{author}{\bibinfo{person}{Andrew~L. Maas}, \bibinfo{person}{Raymond~E. Daly}, \bibinfo{person}{Peter~T. Pham}, \bibinfo{person}{Dan Huang}, \bibinfo{person}{Andrew~Y. Ng}, {and} \bibinfo{person}{Christopher Potts}.} \bibinfo{year}{2011}\natexlab{}.
\newblock \showarticletitle{Learning Word Vectors for Sentiment Analysis}. In \bibinfo{booktitle}{\emph{Proceedings of the 49th Annual Meeting of the Association for Computational Linguistics: Human Language Technologies}}, \bibfield{editor}{\bibinfo{person}{Dekang Lin}, \bibinfo{person}{Yuji Matsumoto}, {and} \bibinfo{person}{Rada Mihalcea}} (Eds.). \bibinfo{publisher}{Association for Computational Linguistics}, \bibinfo{address}{Portland, Oregon, USA}, \bibinfo{pages}{142--150}.
\newblock
\urldef\tempurl%
\url{https://aclanthology.org/P11-1015}
\showURL{%
\tempurl}


\bibitem[Ng et~al\mbox{.}(2023)]%
        {google-research-identify-contrails-reduce-global-warming}
\bibfield{author}{\bibinfo{person}{Joe Ng}, \bibinfo{person}{Carl Elkin}, \bibinfo{person}{Aaron Sarna}, \bibinfo{person}{Walter Reade}, {and} \bibinfo{person}{Maggie Demkin}.} \bibinfo{year}{2023}\natexlab{}.
\newblock \bibinfo{title}{Google Research - Identify Contrails to Reduce Global Warming}.
\newblock
\newblock
\urldef\tempurl%
\url{https://kaggle.com/competitions/google-research-identify-contrails-reduce-global-warming}
\showURL{%
\tempurl}


\bibitem[Park et~al\mbox{.}(2023)]%
        {park2023generative}
\bibfield{author}{\bibinfo{person}{Joon~Sung Park}, \bibinfo{person}{Joseph~C. O'Brien}, \bibinfo{person}{Carrie~J. Cai}, \bibinfo{person}{Meredith~Ringel Morris}, \bibinfo{person}{Percy Liang}, {and} \bibinfo{person}{Michael~S. Bernstein}.} \bibinfo{year}{2023}\natexlab{}.
\newblock \bibinfo{title}{Generative Agents: Interactive Simulacra of Human Behavior}.
\newblock
\newblock
\showeprint[arxiv]{2304.03442}~[cs.HC]


\bibitem[Qian et~al\mbox{.}(2023)]%
        {qian2023communicative}
\bibfield{author}{\bibinfo{person}{Chen Qian}, \bibinfo{person}{Xin Cong}, \bibinfo{person}{Wei Liu}, \bibinfo{person}{Cheng Yang}, \bibinfo{person}{Weize Chen}, \bibinfo{person}{Yusheng Su}, \bibinfo{person}{Yufan Dang}, \bibinfo{person}{Jiahao Li}, \bibinfo{person}{Juyuan Xu}, \bibinfo{person}{Dahai Li}, \bibinfo{person}{Zhiyuan Liu}, {and} \bibinfo{person}{Maosong Sun}.} \bibinfo{year}{2023}\natexlab{}.
\newblock \bibinfo{title}{Communicative Agents for Software Development}.
\newblock
\newblock
\showeprint[arxiv]{2307.07924}~[cs.SE]


\bibitem[Rozière et~al\mbox{.}(2024)]%
        {rozière2024code}
\bibfield{author}{\bibinfo{person}{Baptiste Rozière}, \bibinfo{person}{Jonas Gehring}, \bibinfo{person}{Fabian Gloeckle}, \bibinfo{person}{Sten Sootla}, \bibinfo{person}{Itai Gat}, \bibinfo{person}{Xiaoqing~Ellen Tan}, \bibinfo{person}{Yossi Adi}, \bibinfo{person}{Jingyu Liu}, \bibinfo{person}{Romain Sauvestre}, \bibinfo{person}{Tal Remez}, \bibinfo{person}{Jérémy Rapin}, \bibinfo{person}{Artyom Kozhevnikov}, \bibinfo{person}{Ivan Evtimov}, \bibinfo{person}{Joanna Bitton}, \bibinfo{person}{Manish Bhatt}, \bibinfo{person}{Cristian~Canton Ferrer}, \bibinfo{person}{Aaron Grattafiori}, \bibinfo{person}{Wenhan Xiong}, \bibinfo{person}{Alexandre Défossez}, \bibinfo{person}{Jade Copet}, \bibinfo{person}{Faisal Azhar}, \bibinfo{person}{Hugo Touvron}, \bibinfo{person}{Louis Martin}, \bibinfo{person}{Nicolas Usunier}, \bibinfo{person}{Thomas Scialom}, {and} \bibinfo{person}{Gabriel Synnaeve}.} \bibinfo{year}{2024}\natexlab{}.
\newblock \bibinfo{title}{Code Llama: Open Foundation Models for Code}.
\newblock
\newblock
\showeprint[arxiv]{2308.12950}~[cs.CL]


\bibitem[Salehin et~al\mbox{.}(2024)]%
        {SALEHIN202452}
\bibfield{author}{\bibinfo{person}{Imrus Salehin}, \bibinfo{person}{Md.~Shamiul Islam}, \bibinfo{person}{Pritom Saha}, \bibinfo{person}{S.M. Noman}, \bibinfo{person}{Azra Tuni}, \bibinfo{person}{Md.~Mehedi Hasan}, {and} \bibinfo{person}{Md.~Abu Baten}.} \bibinfo{year}{2024}\natexlab{}.
\newblock \showarticletitle{AutoML: A systematic review on automated machine learning with neural architecture search}.
\newblock \bibinfo{journal}{\emph{Journal of Information and Intelligence}} \bibinfo{volume}{2}, \bibinfo{number}{1} (\bibinfo{year}{2024}), \bibinfo{pages}{52--81}.
\newblock
\showISSN{2949-7159}
\urldef\tempurl%
\url{https://doi.org/10.1016/j.jiixd.2023.10.002}
\showDOI{\tempurl}


\bibitem[Shen et~al\mbox{.}(2024)]%
        {shen2024small}
\bibfield{author}{\bibinfo{person}{Weizhou Shen}, \bibinfo{person}{Chenliang Li}, \bibinfo{person}{Hongzhan Chen}, \bibinfo{person}{Ming Yan}, \bibinfo{person}{Xiaojun Quan}, \bibinfo{person}{Hehong Chen}, \bibinfo{person}{Ji Zhang}, {and} \bibinfo{person}{Fei Huang}.} \bibinfo{year}{2024}\natexlab{}.
\newblock \bibinfo{title}{Small LLMs Are Weak Tool Learners: A Multi-LLM Agent}.
\newblock
\newblock
\showeprint[arxiv]{2401.07324}~[cs.AI]


\bibitem[Tang et~al\mbox{.}(2024)]%
        {tang2024mlbenchevaluatinglargelanguage}
\bibfield{author}{\bibinfo{person}{Xiangru Tang}, \bibinfo{person}{Yuliang Liu}, \bibinfo{person}{Zefan Cai}, \bibinfo{person}{Yanjun Shao}, \bibinfo{person}{Junjie Lu}, \bibinfo{person}{Yichi Zhang}, \bibinfo{person}{Zexuan Deng}, \bibinfo{person}{Helan Hu}, \bibinfo{person}{Kaikai An}, \bibinfo{person}{Ruijun Huang}, \bibinfo{person}{Shuzheng Si}, \bibinfo{person}{Sheng Chen}, \bibinfo{person}{Haozhe Zhao}, \bibinfo{person}{Liang Chen}, \bibinfo{person}{Yan Wang}, \bibinfo{person}{Tianyu Liu}, \bibinfo{person}{Zhiwei Jiang}, \bibinfo{person}{Baobao Chang}, \bibinfo{person}{Yin Fang}, \bibinfo{person}{Yujia Qin}, \bibinfo{person}{Wangchunshu Zhou}, \bibinfo{person}{Yilun Zhao}, \bibinfo{person}{Arman Cohan}, {and} \bibinfo{person}{Mark Gerstein}.} \bibinfo{year}{2024}\natexlab{}.
\newblock \bibinfo{title}{ML-Bench: Evaluating Large Language Models and Agents for Machine Learning Tasks on Repository-Level Code}.
\newblock
\newblock
\showeprint[arxiv]{2311.09835}~[cs.CL]
\urldef\tempurl%
\url{https://arxiv.org/abs/2311.09835}
\showURL{%
\tempurl}


\bibitem[Team et~al\mbox{.}(2023)]%
        {geminiteam2023gemini}
\bibfield{author}{\bibinfo{person}{Gemini Team}, \bibinfo{person}{Rohan Anil}, \bibinfo{person}{Sebastian Borgeaud}, {et~al\mbox{.}}} \bibinfo{year}{2023}\natexlab{}.
\newblock \bibinfo{title}{Gemini: A Family of Highly Capable Multimodal Models}.
\newblock
\newblock
\showeprint[arxiv]{2312.11805}~[cs.CL]


\bibitem[Touvron et~al\mbox{.}(2023)]%
        {touvron2023llama}
\bibfield{author}{\bibinfo{person}{Hugo Touvron}, \bibinfo{person}{Louis Martin}, \bibinfo{person}{Kevin Stone}, \bibinfo{person}{Peter Albert}, \bibinfo{person}{Amjad Almahairi}, \bibinfo{person}{Yasmine Babaei}, \bibinfo{person}{Nikolay Bashlykov}, \bibinfo{person}{Soumya Batra}, \bibinfo{person}{Prajjwal Bhargava}, \bibinfo{person}{Shruti Bhosale}, \bibinfo{person}{Dan Bikel}, \bibinfo{person}{Lukas Blecher}, \bibinfo{person}{Cristian~Canton Ferrer}, \bibinfo{person}{Moya Chen}, \bibinfo{person}{Guillem Cucurull}, \bibinfo{person}{David Esiobu}, \bibinfo{person}{Jude Fernandes}, \bibinfo{person}{Jeremy Fu}, \bibinfo{person}{Wenyin Fu}, \bibinfo{person}{Brian Fuller}, \bibinfo{person}{Cynthia Gao}, \bibinfo{person}{Vedanuj Goswami}, \bibinfo{person}{Naman Goyal}, \bibinfo{person}{Anthony Hartshorn}, \bibinfo{person}{Saghar Hosseini}, \bibinfo{person}{Rui Hou}, \bibinfo{person}{Hakan Inan}, \bibinfo{person}{Marcin Kardas}, \bibinfo{person}{Viktor Kerkez}, \bibinfo{person}{Madian Khabsa},
  \bibinfo{person}{Isabel Kloumann}, \bibinfo{person}{Artem Korenev}, \bibinfo{person}{Punit~Singh Koura}, \bibinfo{person}{Marie-Anne Lachaux}, \bibinfo{person}{Thibaut Lavril}, \bibinfo{person}{Jenya Lee}, \bibinfo{person}{Diana Liskovich}, \bibinfo{person}{Yinghai Lu}, \bibinfo{person}{Yuning Mao}, \bibinfo{person}{Xavier Martinet}, \bibinfo{person}{Todor Mihaylov}, \bibinfo{person}{Pushkar Mishra}, \bibinfo{person}{Igor Molybog}, \bibinfo{person}{Yixin Nie}, \bibinfo{person}{Andrew Poulton}, \bibinfo{person}{Jeremy Reizenstein}, \bibinfo{person}{Rashi Rungta}, \bibinfo{person}{Kalyan Saladi}, \bibinfo{person}{Alan Schelten}, \bibinfo{person}{Ruan Silva}, \bibinfo{person}{Eric~Michael Smith}, \bibinfo{person}{Ranjan Subramanian}, \bibinfo{person}{Xiaoqing~Ellen Tan}, \bibinfo{person}{Binh Tang}, \bibinfo{person}{Ross Taylor}, \bibinfo{person}{Adina Williams}, \bibinfo{person}{Jian~Xiang Kuan}, \bibinfo{person}{Puxin Xu}, \bibinfo{person}{Zheng Yan}, \bibinfo{person}{Iliyan Zarov}, \bibinfo{person}{Yuchen
  Zhang}, \bibinfo{person}{Angela Fan}, \bibinfo{person}{Melanie Kambadur}, \bibinfo{person}{Sharan Narang}, \bibinfo{person}{Aurelien Rodriguez}, \bibinfo{person}{Robert Stojnic}, \bibinfo{person}{Sergey Edunov}, {and} \bibinfo{person}{Thomas Scialom}.} \bibinfo{year}{2023}\natexlab{}.
\newblock \bibinfo{title}{Llama 2: Open Foundation and Fine-Tuned Chat Models}.
\newblock
\newblock
\showeprint[arxiv]{2307.09288}~[cs.CL]


\bibitem[Veličković et~al\mbox{.}(2022)]%
        {veličković2022clrs}
\bibfield{author}{\bibinfo{person}{Petar Veličković}, \bibinfo{person}{Adrià~Puigdomènech Badia}, \bibinfo{person}{David Budden}, \bibinfo{person}{Razvan Pascanu}, \bibinfo{person}{Andrea Banino}, \bibinfo{person}{Misha Dashevskiy}, \bibinfo{person}{Raia Hadsell}, {and} \bibinfo{person}{Charles Blundell}.} \bibinfo{year}{2022}\natexlab{}.
\newblock \bibinfo{title}{The CLRS Algorithmic Reasoning Benchmark}.
\newblock
\newblock
\showeprint[arxiv]{2205.15659}~[cs.LG]


\bibitem[Wang et~al\mbox{.}(2023)]%
        {wang2023fusing}
\bibfield{author}{\bibinfo{person}{Hongyi Wang}, \bibinfo{person}{Felipe~Maia Polo}, \bibinfo{person}{Yuekai Sun}, \bibinfo{person}{Souvik Kundu}, \bibinfo{person}{Eric Xing}, {and} \bibinfo{person}{Mikhail Yurochkin}.} \bibinfo{year}{2023}\natexlab{}.
\newblock \bibinfo{title}{Fusing Models with Complementary Expertise}.
\newblock
\newblock
\showeprint[arxiv]{2310.01542}~[cs.LG]


\bibitem[Wang et~al\mbox{.}(2024)]%
        {wang2024executable}
\bibfield{author}{\bibinfo{person}{Xingyao Wang}, \bibinfo{person}{Yangyi Chen}, \bibinfo{person}{Lifan Yuan}, \bibinfo{person}{Yizhe Zhang}, \bibinfo{person}{Yunzhu Li}, \bibinfo{person}{Hao Peng}, {and} \bibinfo{person}{Heng Ji}.} \bibinfo{year}{2024}\natexlab{}.
\newblock \bibinfo{title}{Executable Code Actions Elicit Better LLM Agents}.
\newblock
\newblock
\showeprint[arxiv]{2402.01030}~[cs.CL]


\bibitem[Warstadt et~al\mbox{.}(2023)]%
        {warstadt2023papers}
\bibfield{author}{\bibinfo{person}{Alex Warstadt}, \bibinfo{person}{Leshem Choshen}, \bibinfo{person}{Aaron Mueller}, \bibinfo{person}{Adina Williams}, \bibinfo{person}{Ethan Wilcox}, {and} \bibinfo{person}{Chengxu Zhuang}.} \bibinfo{year}{2023}\natexlab{}.
\newblock \bibinfo{title}{Call for Papers -- The BabyLM Challenge: Sample-efficient pretraining on a developmentally plausible corpus}.
\newblock
\newblock
\showeprint[arxiv]{2301.11796}~[cs.CL]


\bibitem[Woodward et~al\mbox{.}(2023)]%
        {fathomnet-out-of-sample-detection}
\bibfield{author}{\bibinfo{person}{Ben Woodward}, \bibinfo{person}{eor123}, \bibinfo{person}{Genevieve Patterson}, {and} \bibinfo{person}{Lilli Carlsen}.} \bibinfo{year}{2023}\natexlab{}.
\newblock \bibinfo{title}{FathomNet 2023}.
\newblock
\newblock
\urldef\tempurl%
\url{https://kaggle.com/competitions/fathomnet-out-of-sample-detection}
\showURL{%
\tempurl}


\bibitem[Wu et~al\mbox{.}(2023)]%
        {wu2023autogen}
\bibfield{author}{\bibinfo{person}{Qingyun Wu}, \bibinfo{person}{Gagan Bansal}, \bibinfo{person}{Jieyu Zhang}, \bibinfo{person}{Yiran Wu}, \bibinfo{person}{Beibin Li}, \bibinfo{person}{Erkang Zhu}, \bibinfo{person}{Li Jiang}, \bibinfo{person}{Xiaoyun Zhang}, \bibinfo{person}{Shaokun Zhang}, \bibinfo{person}{Jiale Liu}, \bibinfo{person}{Ahmed~Hassan Awadallah}, \bibinfo{person}{Ryen~W White}, \bibinfo{person}{Doug Burger}, {and} \bibinfo{person}{Chi Wang}.} \bibinfo{year}{2023}\natexlab{}.
\newblock \bibinfo{title}{AutoGen: Enabling Next-Gen LLM Applications via Multi-Agent Conversation}.
\newblock
\newblock
\showeprint[arxiv]{2308.08155}~[cs.AI]


\bibitem[Yeadon et~al\mbox{.}(2024)]%
        {yeadon2024comparison}
\bibfield{author}{\bibinfo{person}{Will Yeadon}, \bibinfo{person}{Alex Peach}, {and} \bibinfo{person}{Craig~P. Testrow}.} \bibinfo{year}{2024}\natexlab{}.
\newblock \bibinfo{title}{A comparison of Human, GPT-3.5, and GPT-4 Performance in a University-Level Coding Course}.
\newblock
\newblock
\showeprint[arxiv]{2403.16977}~[cs.CL]


\bibitem[Yi et~al\mbox{.}(2024)]%
        {yi2024survey}
\bibfield{author}{\bibinfo{person}{Zihao Yi}, \bibinfo{person}{Jiarui Ouyang}, \bibinfo{person}{Yuwen Liu}, \bibinfo{person}{Tianhao Liao}, \bibinfo{person}{Zhe Xu}, {and} \bibinfo{person}{Ying Shen}.} \bibinfo{year}{2024}\natexlab{}.
\newblock \bibinfo{title}{A Survey on Recent Advances in LLM-Based Multi-turn Dialogue Systems}.
\newblock
\newblock
\showeprint[arxiv]{2402.18013}~[cs.CL]


\bibitem[Yue et~al\mbox{.}(2024)]%
        {yue2024large}
\bibfield{author}{\bibinfo{person}{Murong Yue}, \bibinfo{person}{Jie Zhao}, \bibinfo{person}{Min Zhang}, \bibinfo{person}{Liang Du}, {and} \bibinfo{person}{Ziyu Yao}.} \bibinfo{year}{2024}\natexlab{}.
\newblock \bibinfo{title}{Large Language Model Cascades with Mixture of Thoughts Representations for Cost-efficient Reasoning}.
\newblock
\newblock
\showeprint[arxiv]{2310.03094}~[cs.CL]


\bibitem[Zan et~al\mbox{.}(2023)]%
        {zan-etal-2023-large}
\bibfield{author}{\bibinfo{person}{Daoguang Zan}, \bibinfo{person}{Bei Chen}, \bibinfo{person}{Fengji Zhang}, \bibinfo{person}{Dianjie Lu}, \bibinfo{person}{Bingchao Wu}, \bibinfo{person}{Bei Guan}, \bibinfo{person}{Wang Yongji}, {and} \bibinfo{person}{Jian-Guang Lou}.} \bibinfo{year}{2023}\natexlab{}.
\newblock \showarticletitle{Large Language Models Meet {NL}2{C}ode: A Survey}. In \bibinfo{booktitle}{\emph{Proceedings of the 61st Annual Meeting of the Association for Computational Linguistics (Volume 1: Long Papers)}}, \bibfield{editor}{\bibinfo{person}{Anna Rogers}, \bibinfo{person}{Jordan Boyd-Graber}, {and} \bibinfo{person}{Naoaki Okazaki}} (Eds.). \bibinfo{publisher}{Association for Computational Linguistics}, \bibinfo{address}{Toronto, Canada}, \bibinfo{pages}{7443--7464}.
\newblock
\urldef\tempurl%
\url{https://doi.org/10.18653/v1/2023.acl-long.411}
\showDOI{\tempurl}


\bibitem[Zan et~al\mbox{.}(2024)]%
        {zan2024codesnaturallanguagecode}
\bibfield{author}{\bibinfo{person}{Daoguang Zan}, \bibinfo{person}{Ailun Yu}, \bibinfo{person}{Wei Liu}, \bibinfo{person}{Dong Chen}, \bibinfo{person}{Bo Shen}, \bibinfo{person}{Wei Li}, \bibinfo{person}{Yafen Yao}, \bibinfo{person}{Yongshun Gong}, \bibinfo{person}{Xiaolin Chen}, \bibinfo{person}{Bei Guan}, \bibinfo{person}{Zhiguang Yang}, \bibinfo{person}{Yongji Wang}, \bibinfo{person}{Qianxiang Wang}, {and} \bibinfo{person}{Lizhen Cui}.} \bibinfo{year}{2024}\natexlab{}.
\newblock \bibinfo{title}{CodeS: Natural Language to Code Repository via Multi-Layer Sketch}.
\newblock
\newblock
\showeprint[arxiv]{2403.16443}~[cs.CL]
\urldef\tempurl%
\url{https://arxiv.org/abs/2403.16443}
\showURL{%
\tempurl}


\bibitem[Zhang et~al\mbox{.}(2023)]%
        {zhang2023ecoassistant}
\bibfield{author}{\bibinfo{person}{Jieyu Zhang}, \bibinfo{person}{Ranjay Krishna}, \bibinfo{person}{Ahmed~H. Awadallah}, {and} \bibinfo{person}{Chi Wang}.} \bibinfo{year}{2023}\natexlab{}.
\newblock \bibinfo{title}{EcoAssistant: Using LLM Assistant More Affordably and Accurately}.
\newblock
\newblock
\showeprint[arxiv]{2310.03046}~[cs.SE]


\bibitem[Zhang et~al\mbox{.}(2024b)]%
        {zhang2024codeagent}
\bibfield{author}{\bibinfo{person}{Kechi Zhang}, \bibinfo{person}{Jia Li}, \bibinfo{person}{Ge Li}, \bibinfo{person}{Xianjie Shi}, {and} \bibinfo{person}{Zhi Jin}.} \bibinfo{year}{2024}\natexlab{b}.
\newblock \bibinfo{title}{CodeAgent: Enhancing Code Generation with Tool-Integrated Agent Systems for Real-World Repo-level Coding Challenges}.
\newblock
\newblock
\showeprint[arxiv]{2401.07339}~[cs.SE]


\bibitem[Zhang et~al\mbox{.}(2024c)]%
        {zhang2024mlcopilotunleashingpowerlarge}
\bibfield{author}{\bibinfo{person}{Lei Zhang}, \bibinfo{person}{Yuge Zhang}, \bibinfo{person}{Kan Ren}, \bibinfo{person}{Dongsheng Li}, {and} \bibinfo{person}{Yuqing Yang}.} \bibinfo{year}{2024}\natexlab{c}.
\newblock \bibinfo{title}{MLCopilot: Unleashing the Power of Large Language Models in Solving Machine Learning Tasks}.
\newblock
\newblock
\showeprint[arxiv]{2304.14979}~[cs.LG]
\urldef\tempurl%
\url{https://arxiv.org/abs/2304.14979}
\showURL{%
\tempurl}


\bibitem[Zhang et~al\mbox{.}(2024a)]%
        {zhang2024unifying}
\bibfield{author}{\bibinfo{person}{Ziyin Zhang}, \bibinfo{person}{Chaoyu Chen}, \bibinfo{person}{Bingchang Liu}, \bibinfo{person}{Cong Liao}, \bibinfo{person}{Zi Gong}, \bibinfo{person}{Hang Yu}, \bibinfo{person}{Jianguo Li}, {and} \bibinfo{person}{Rui Wang}.} \bibinfo{year}{2024}\natexlab{a}.
\newblock \bibinfo{title}{Unifying the Perspectives of NLP and Software Engineering: A Survey on Language Models for Code}.
\newblock
\newblock
\showeprint[arxiv]{2311.07989}~[cs.CL]


\bibitem[Zheng et~al\mbox{.}(2024)]%
        {zheng2024survey}
\bibfield{author}{\bibinfo{person}{Zibin Zheng}, \bibinfo{person}{Kaiwen Ning}, \bibinfo{person}{Yanlin Wang}, \bibinfo{person}{Jingwen Zhang}, \bibinfo{person}{Dewu Zheng}, \bibinfo{person}{Mingxi Ye}, {and} \bibinfo{person}{Jiachi Chen}.} \bibinfo{year}{2024}\natexlab{}.
\newblock \bibinfo{title}{A Survey of Large Language Models for Code: Evolution, Benchmarking, and Future Trends}.
\newblock
\newblock
\showeprint[arxiv]{2311.10372}~[cs.SE]


\bibitem[Zhong et~al\mbox{.}(2024)]%
        {zhong2024ldb}
\bibfield{author}{\bibinfo{person}{Lily Zhong}, \bibinfo{person}{Zilong Wang}, {and} \bibinfo{person}{Jingbo Shang}.} \bibinfo{year}{2024}\natexlab{}.
\newblock \bibinfo{title}{LDB: A Large Language Model Debugger via Verifying Runtime Execution Step-by-step}.
\newblock
\newblock
\showeprint[arxiv]{2402.16906}~[cs.SE]


\bibitem[Zhu et~al\mbox{.}(2023)]%
        {zhu2023multilingual}
\bibfield{author}{\bibinfo{person}{Wenhao Zhu}, \bibinfo{person}{Hongyi Liu}, \bibinfo{person}{Qingxiu Dong}, \bibinfo{person}{Jingjing Xu}, \bibinfo{person}{Shujian Huang}, \bibinfo{person}{Lingpeng Kong}, \bibinfo{person}{Jiajun Chen}, {and} \bibinfo{person}{Lei Li}.} \bibinfo{year}{2023}\natexlab{}.
\newblock \bibinfo{title}{Multilingual Machine Translation with Large Language Models: Empirical Results and Analysis}.
\newblock
\newblock
\showeprint[arxiv]{2304.04675}~[cs.CL]


\end{thebibliography}

\end{document}